\documentclass{elsart}

 \usepackage{graphics}
 \usepackage{graphicx}
 \usepackage{epsfig}
\usepackage[latin1]{inputenc}
\usepackage{amssymb}
\usepackage{amsmath,amsbsy,amscd,amstext}
\newcommand{\dept}{\left(t\right)}
\newcommand{\depr}{\left(\mathbf{r}\right)}

\newcommand{\depo}{\left(0\right)}
\newcommand{\deprt}{\left(\mathbf{r},t\right)}
\newcommand{\deprpt}{\left(\mathbf{r}^{\prime},t\right)}

\begin{document}

\begin{frontmatter}

% Title, authors and addresses

% use the thanksref command within \title, \author or \address for footnotes;
% use the corauthref command within \author for corresponding author footnotes;
% use the ead command for the email address,
% and the form \ead[url] for the home page:
 \title{Ramsey fringes formation during excitation of topological modes in a Bose-Einstein condensate}
% \thanks[label1]{}
 \author[usp]{E. R. F. Ramos},
 \ead{edmir@ursa.ifsc.usp.br}
\author[usp]{L. Sanz}, 
\author[jinr]{V. I. Yukalov} and
\author[usp]{V. S. Bagnato}
% \ead[url]{home page}
% \thanks[label2]{}
% \corauth[cor1]{}
 \address[usp]{Instituto de Física de São Carlos, Universidade de
São Paulo,\\Caixa Postal 369, 13560-970, São Carlos-SP Brazil}
 \address[jinr]{Bogolubov Laboratory of Theoretical Physics, Joint Institute for Nuclear Research
Dubna 141980, Russia}
% \thanks[label3]{}
% use optional labels to link authors explicitly to addresses:
% \author[label1,label2]{}
% \address[label1]{}
% \address[label2]{}

\author{}

\address{}

\begin{abstract}
% Text of abstract
The Ramsey fringes formation during the excitation of topological  
coherent modes of a Bose-Einstein condensate by an external 
modulating field is considered. The Ramsey fringes appear when a series 
of pulses of the excitation field is applied. In both Rabi and Ramsey 
interrogations, there is a shift of the population maximum transfer due  
to the strong non-linearity present in the system. It is found that the  
Ramsey pattern itself retains information about the accumulated relative  
phase between both ground and excited coherent modes.
\end{abstract}

\end{frontmatter}

% main text
\section{Introduction}
\label{sec:intro}

The creation of Bose-Einstein condensates (BEC) in non-ground states, 
as originally proposed by Yukalov, Yukalova, and Bagnato 
~\cite{Yukalov97,ReviewVSB01} is nowadays a topic of wide interest. 
This method allows for the direct formation of fragmented 
nonequilibrium condensates (see the review article by Leggett
~\cite{LeggettRMP}). One of important applications of coupling between 
different coherent modes of BEC is the possibility to produce various  
transverse modes in the atom laser 
~\cite{Mewes97,NistBoser,MunichBoser,YaleBoser}. Many experiments are  
devoted to the study of properties of trapped BECs by coupling their  
collective states~\cite{Myatt97,Matthews98}.

Starting with a sample of atoms Bose-condensed in the ground state
of a confining potential, it is possible to promote atoms from one
trap level to another by using \emph{resonance
excitation}~\cite{Yukalov97,Yukalov02}. This is done by applying an  
additional weak external field, with a fixed spatial distribution, and 
oscillating in time with a frequency near the transition frequency  
between the ground and an excited state. Previous calculations 
demonstrated the possibility of macroscopic transfer between the levels, 
confirming the feasibility of this procedure, and analyzed several  
applications~\cite{Yukalov97,ReviewVSB01,Yukalov02}.

In the present work, we investigate the formation of the Ramsey-like  
fringes, due to the interference between the ground and a non-ground 
states of BEC, excited by means of a near resonant field. We keep in mind  
a time-domain version of the separated oscillatory-field method, as 
developed by Ramsey~\cite{Ramsey49,Ramsey50}, consisting of a sequence of 
two Rabi $\pi/2$-pulses, which are equivalent to the oscillatory fields of 
the Ramsey method. Previously, the formation of Ramsey fringes in double 
Bose-Einstein condensates~\cite{Eschmann99} was studied. But this is for  
the first time that the Ramsey patterns are obtained, when BECs states  
with different quantum numbers, associated with the trap potential, are  
coupled. The eventual measurement of those fringes would quantify the 
coherence of the process. An experimental setup for the observation of the 
topological coherent modes, through spatial distribution observation, is  
presently in progress in our research group~\cite{Magalhaes05,Henn06}.

This paper is structured as follows. In Sec.~\ref{sec:Gross}, we briefly 
review the dynamics of the coherent modes, based on the Gross-Pitaevskii  
equation for BEC~\cite{Dalfovo99}. Also, we recall the idea of the  
resonant excitation, which makes it possible to couple the ground and  
non-ground states. Section~\ref{sec:Ramsey} is devoted to the formation  
of the Ramsey-like fringes and to the possibility of their experimental  
observation. In Section~\ref{sec:conclusions}, we summarize our results.

\section{Gross-Pitaevskii equation and coherent modes}

\label{sec:Gross} At low temperatures, dilute Bose gas, as is known, is  
well described by the Gross-Pitaevskii equation. This equation 
describes the coherent states of the Bose system with the Hamiltonian 
\begin{eqnarray}
\label{eq:Hmany}
\hat{H}&=&\int\hat{\Psi}^{\dagger}\deprt\left[-\frac{\hbar^2\nabla^2}{2m_0}
+V_{\mbox{ho}}\deprt\right]\hat{\Psi}\deprt d\mathbf{r}+\nonumber\\
&&+\frac{1}{2}\int
d\mathbf{r}d\mathbf{r}^{\prime}\hat{\Psi}^{\dagger}\deprt\hat{\Psi}^{\dagger}
\deprpt
V\left(\mathbf{r}-\mathbf{r}^{\prime}\right)\hat{\Psi}\deprpt\hat{\Psi}\deprt,
\end{eqnarray}
with  $N$ bosons assumed to be confined by a harmonic trap potential, 
$V_{\mbox{ho}}\deprt$. The $\hat{\Psi}\deprt$
($\hat{\Psi}^{\dagger}\deprt$) are the boson field operators that
annihilate (create) an atom at position $\mathbf{r}$ and
$V\left(\mathbf{r}-\mathbf{r}^{\prime}\right)$ is a two-body
interaction due to atomic collisions. In a dilute cold gas, the most
relevant scattering process is associated with elastic binary
collisions at low energy~\cite{LeggettRMP,Dalfovo99}, giving the effective 
interaction potential
\begin{equation}
V\left(\mathbf{r}-\mathbf{r}^{\prime}\right)=A_s\delta\left(\mathbf{r}-\mathbf{r}^{\prime}\right)
=(N-1)\frac{4\pi\hbar^2a_s}{m_0}\delta\left(\mathbf{r}-\mathbf{r}^{\prime}\right).
\label{eq:Vr}
\end{equation}
Here, $a_s$ is the ``zero-energy'' s-\textit{wave scattering length}. For 
an effective attractive (repulsive) interaction $a_s$ is negative 
(positive).  The confining harmonic potential is written as
\begin{equation}
V_{\mbox{ho}}\deprt=\frac{m_0}{2}\left(\omega_x^2x^2+\omega_y^2y^2+
\omega_z^2z^2\right)
\label{eq:Vtrap}
\end{equation}
where $m_0$ is the atomic mass and $\omega_k$ ($k=x,y,z$) are the
trap oscillation frequencies along each axis. 

The topological coherent modes are the solutions to the stationary 
Gross-Pitaevskii equation. The nonlinear coherent modes form an     
overcomplete basis and are normalized ($\left|\varphi_j\right|^2=1$). The  
eigenvalue problem for stationary functions is given by~\cite{Yukalov02}
\begin{equation}
\hat{H}\left[\varphi_{j}\right]\varphi_{j}=E_{j}\varphi_{j}
\label{eq:eigen}
\end{equation}
where
\begin{equation}
\hat{H}\left[\varphi_{j}\right]=-\frac{\hbar^2\nabla^2}{2m_0}+
V_{\mbox{ho}}\deprt
+A_s\left|\varphi_{j}\right|^2
\label{eq:Hvarphi}
\end{equation}
follows from the Hamiltonian (\ref{eq:Hmany}).

The resonant pumping makes it possible a coherent transfer of
condensed atoms between the collective levels of atoms in the harmonic 
trap, as is described in Ref.~\cite{Yukalov02}. At initial time, $N$  
condensed atoms are assumed to be in the ground state of the trap with a   
frequency $\omega_0$. The aim is to transfer the atoms to a non-ground  
state (labelled as $p$). The energy difference of the level-$p$, relative 
to the ground state, is $\hbar\left(\omega_p-\omega_0\right)$. Following 
Yukalov {\it et al.}~\cite{Yukalov02}, this transfer can be obtained  
through the action of an external oscillatory field given by
\begin{equation}
V_p\deprt=V\left(\mathbf{r}\right)\cos{\omega t} \; .
\label{eq:Vp}
\end{equation}

The corresponding non-linear Schrödinger equation, associated with the  
Hamiltonian (\ref{eq:Hvarphi}), writes as
\begin{equation}
i\hbar\frac{\partial\phi\deprt}{\partial
t}=\left[\hat{H}+V_{p}\deprt\right]\phi\deprt. \label{eq:NLSE}
\end{equation}
The modulating field is called external in the sense that it is not a  
part of the harmonic trap setup, but rather the field $V_p\deprt$ is a  
perturbation in Eq.(\ref{eq:NLSE}). The solution $\phi\deprt$, can be  
expressed as a sum over the coherent modes: $\phi\deprt=\sum_n
c_n\dept\varphi_n\deprt$. Although one could consider a more general
form of the external field for transferring atoms between an arbitrary 
pair of collective levels~\cite{Yukalov02}, it is sufficient to work 
with the potential (\ref{eq:Vp}) in order to create the Ramsey 
interference fringes.

The functions $c_0\dept$, associated with ground state, and $c_p\dept$,  
ascribed to an excited state, determine the behavior of the populations in  
each mode 
\begin{equation}
n_0\dept=\left|c_0\dept\right|^2\; , \qquad
n_p\dept=\left|c_p\dept\right|^2.
\end{equation}
There are some conditions on the involved physical parameters in
order to observe a macroscopic population of a non-ground state
($n_p\neq 0$): First, it is required that the resonance condition 
between the external field of frequency $\omega$ and the frequency
associated with the level transition, $\omega_{p0}=\omega_p-\omega_0$,  
be fulfilled. Also, the detuning $\Delta\omega=\omega-\omega_{p0}$ 
should be small enough, $\left|\Delta\omega/\omega_{p0}\right|\ll 1$. 
At the same time, one should remember that not solely the external  
coupling filed, but also the interatomic collisions cause atomic  
transitions between the modes. In order to quantify both effects, it is 
convenient to define the interaction intensity $\alpha$, associated with  
interatomic collisions, and the transition amplitude $\beta$ of the 
coupling field as
\begin{eqnarray}
\alpha_{m,k}&=&\frac{A_s}{\hbar}\int{\left|\varphi_m\depr\right|^2
\left(2\left|\varphi_k\depr\right|^2-\left|\varphi_m\depr\right|^2\right)
d\mathbf{r}}, \nonumber \\
\beta&=&\frac{1}{\hbar}\int\varphi^*_0\depr V_p\depr\varphi_p\depr
d\mathbf{r}. \label{eq:amplitudes}
\end{eqnarray}
For simplicity, we set $\alpha\equiv\alpha_{0,p}=\alpha_{p,0}$ for the 
folowing analysis. These quantities have dimensions of frequency (Hz). 
Their values are to be smaller than the transition frequency
$\omega_{p0}$, so that 
\begin{eqnarray}
\left|\frac{\alpha}{\omega_{p0}}\right|\ll 1\; ,
\qquad  \left|\frac{\beta}{\omega_{p0}}\right|\ll 1.
\label{eq:conditions}
\end{eqnarray}
For the time dependent process, with the potential $V_p$, the solution  
of Eq.(\ref{eq:NLSE}) can be represented as
\begin{equation}
\phi\deprt=\sum_n c_n\dept\varphi_n\depr\exp^{-\frac{i}{\hbar}E_n t}. 
\label{eq:phi}
\end{equation}
Another condition is that the coefficients $c_n\dept$ evolve slowly
in time, when compared to the oscillatory terms in the equation
(\ref{eq:phi}), so that
\begin{equation}
\frac{\hbar}{E_n}\left|\frac{dc_n}{dt}\right|\ll 1.
\label{eq:modeSeparation}
\end{equation}
It is straightforward to obtain a system of coupled differential
equations for the coefficients $c_{0}$ and $c_p$, whose detailed 
derivation can be found in Ref.~\cite{Yukalov97},
\begin{eqnarray}
i\frac{dc_0}{dt}&=&\alpha n_pc_0+\frac{1}{2}\beta e^{i\Delta\omega t}c_p,
\nonumber\\
i\frac{dc_p}{dt}&=&\alpha n_0c_p+\frac{1}{2}\beta^*e^{-i\Delta\omega
t}c_0. 
\label{eq:system}
\end{eqnarray}
Let us mention that the form of these coupled equations resembles  
the equations of the Rabi two-level problem~\cite{Rabi37}. It is
expected that, if we choose the appropriate physical parameters, Rabi 
oscillations between the topological modes could be observed. The
solution to this system provides us with all necessary information about  
the dynamics of the populations, including both ground and 
non-ground states. An analytical solution of Eqs.(\ref{eq:system})
can be obtained when $\left|\beta\right|\ll \left|\alpha\right|$, as 
in Ref. \cite{ReviewVSB01}. Then the population 
fractions are   
given by
\begin{eqnarray}
n_0&=&1-\frac{\left|\beta\right|^2}{\Omega^2}\sin^2{\frac{\Omega t}{2}}
\nonumber\\
n_p&=&\frac{\left|\beta\right|^2}{\Omega^2}\sin^2{\frac{\Omega
t}{2}}, \label{eq:analiticpop}
\end{eqnarray}
with the effective collective frequency $\Omega$ defined as
\begin{equation}
\Omega=\sqrt{\left|\beta\right|^2+\left[\alpha\left(n_0-n_p\right)-
\Delta\omega\right]^2}.
\label{eq:Omega}
\end{equation}
It is worth emphasizing that this collective frequency is not a constant,  
but a function of the instant populations $n_0$ and $n_p$, and that the  
analytical solution is available under the condition that the populations  
of the modes have small changes. Thus, the population difference $\Delta 
n=\left|n_p-n_0\right|$ is almost a constant. When we are interested in a  
macroscopic population of an excited state, it is necessary to solve the  
coupled equations (\ref{eq:system}) using a numerical procedure based on 
the fourth-order Runge-Kutta method. Thus, we can deal with $\alpha$ and  
$\beta$ values not restricted to the condition $\left|\beta\right|\ll
\left|\alpha\right|$.

%Figure 1
%%%%%%%%%%%%%%%%%%%%%%%%%%%%%%%%%%%%%%%%%%%%%%%%%%%%%%%%%%%%%%%%%%%%%%%%%%%%%%%%%%%%%%%%%%%%%%%%%%
\begin{figure}[ht]
\vspace{0.5cm} \centerline{
\includegraphics[scale=1]{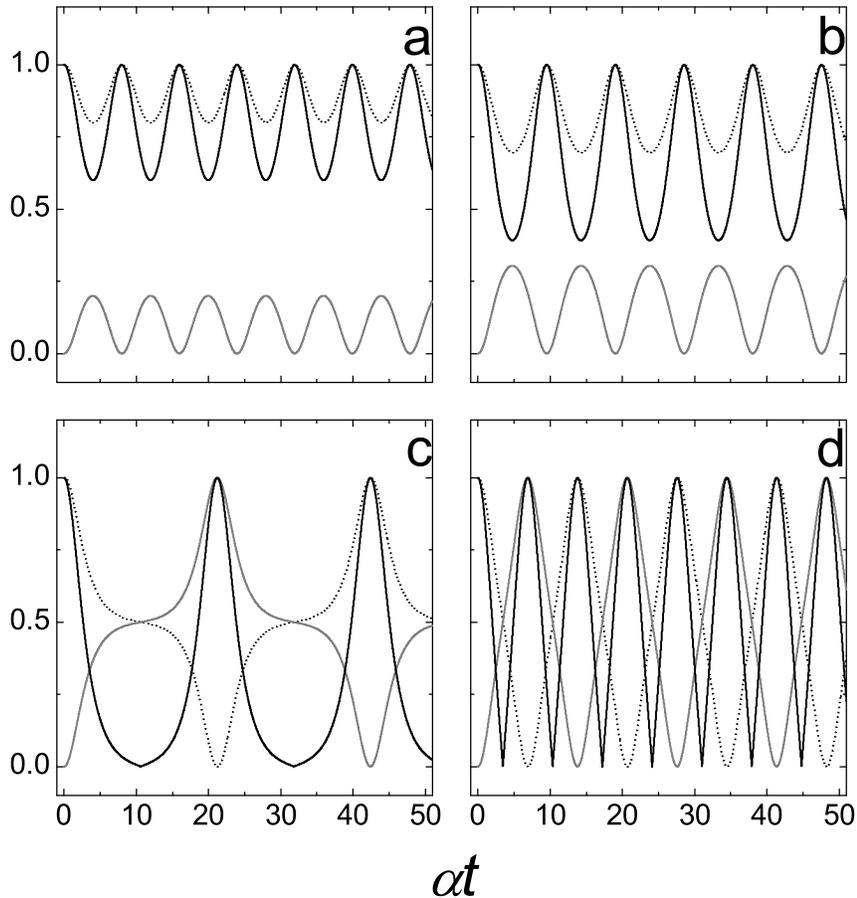}}
\caption{Evolution of the density of populations ($n_0$ in dotted
line, $n_p$ in grey line) and the population unbalance $\Delta
n=\left|n_p-n_0\right|$ (black line) as function of dimensionless
quantity $\alpha t$ for $\Delta\omega\approx 0$ and different values
of coupling parameter $\beta$: (a)~$\beta=0.40\alpha$;
(b)~$\beta=0.46\alpha$; (c)~$\beta\approx 0.50\alpha$ and
(d)~$\beta=0.60\alpha$.} \label{fig:populations}
\end{figure}
%%%%%%%%%%%%%%%%%%%%%%%%%%%%%%%%%%%%%%%%%%%%%%%%%%%%%%%%%%%%%%%%%%%%%%%%%%%%%%%%%%%%%%%%%%%%%%%%%%
In order to find the time dependent solutions for coefficients
$c_0\dept$ and $c_p\dept$, we consider as initial conditions
$c_0\depo=1$ and $c_p\depo=0$ (all $N$ atoms initially condensed in 
the ground state). In figure~\ref{fig:populations}, we plot the
results for the fractional populations, associated with the non-ground 
and excited states and the population difference as functions of the 
dimensionless parameter $\alpha t$. Grey lines show the population
of the non-ground state ($n_p=0$ at initial time) and dotted lines
show $n_0\dept$. The evolution of the population difference, $\Delta
n\dept$, is plotted in black bold lines. For all cases, the detuning is  
fixed so that $\Delta\omega\approx 0$, and we simulate a slight change  
of the external field $V_p$, which is characterized by the coupling  
parameter $\beta$ defined in Eq.(\ref{eq:amplitudes}). From the 
previous works~\cite{Yukalov03a,Yukalov02,ReviewVSB01}, we know that  
there are critical effects associated with the coupled system given by  
Eqs.(\ref{eq:system}), which, however, we shall not consider here.  

Since it is possible to transfer populations between two different  
topological states, under well defined conditions, prescribing when this  
process is efficient, we are in a position to analyze what happens  
if we manipulate the excitation time domain in order to detect Ramsey-like  
interference fringes as the oscillations of the fractional populations.

\section{$\pi/2$ and $\pi$- pulses and the formation of the 
Ramsey-like fringes}
\label{sec:Ramsey}

In this section, we demonstrate the formation of the Ramsey-like 
fringes for the case of two topological modes coupled by an external 
resonant field. Given a certain value of $\beta$, we define the
maximum of the atom population that is transferred from the ground to an   
excited state. It is also possible to estimate the necessary time for the  
coupling to be switched on. When a $\pi$ pulse is applied, the time 
of the coupling, $t_1$, is sufficient for the performance of a complete  
Rabi oscillation. If the coupling is switched off at the time, when the  
half of the maximum population is transferred, the applied pulse is the  
so-called Rabi $\pi/2$-pulse. We show below that Ramsey fringes are  
obtained when two Rabi $\pi/2$-pulses are applied with a time interval  
$\tau$ between them. The latter procedure is similar to that one 
accomplished for the coupled hyperfine levels of Rubidium~\cite{Hall98b},   
and the formation of the fringes confirms the existence of a relative  
phase between both topological modes.

\begin{figure}[ht]
\centerline{\includegraphics[scale=1]{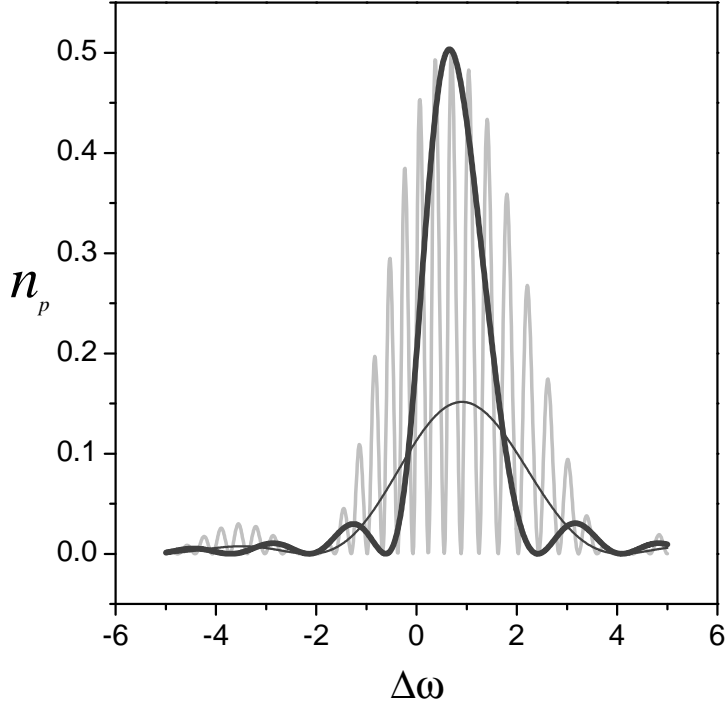}} \caption{The
fractional population of a non-ground state, $n_p$, as a function of 
detuning $\Delta\omega$ for three different kinds of pulse
configurations. For all cases $\beta=0.4\alpha$. Grey line: 
Ramsey fringes due to the application of two $\pi/2$ pulses
with an interval $\tau=8t_1$; Bold black line: a $\pi$
Rabi pulse; Thick black line: $\pi/2$ Rabi pulse.}
\label{fig:ramvsrabi}
\end{figure}

In figure~\ref{fig:ramvsrabi}, we plot our results for the fractional
population density of the ground state, $n_0$, considering three 
pulse configurations, mentioned above, as functions of the detuning 
$\Delta\omega$. The normalized quantity $n_0$ gives us information about  
the atomic population in the mode and an indirect quantification of the  
visibility for the imaging process. The total time of the simulations is  
equal in all three cases. The procedure of obtaining the patterns implies  
the solution of system (\ref{eq:system}) for a fixed value of $\beta$, 
which means a fixed value of the spatial amplitude function in $V_p$,  
allowing for the variation of the detuning $\Delta\omega$.
%Figure 2
%%%%%%%%%%%%%%%%%%%%%%%%%%%%%%%%%%%%%%%%%%%%%%%%%%%%%%%%%%%%%%%%%%%%%%%%%%%%%%%%%%%%%%%%%%%%%%%%%%
%%%%%%%%%%%%%%%%%%%%%%%%%%%%%%%%%%%%%%%%%%%%%%%%%%%%%%%%%%%%%%%%%%%%%%%%%%%%%%%%%%%%%%%%%%%%%%%%%%

In figure~\ref{fig:ramvsrabi}, black lines show the expected behavior of  
$n_0$, when equivalent Rabi pulses are applied. Both, $\pi$ (bold line) 
and $\pi/2$ (thick line), pulses exhibit common features. First, we  
observe a sole peak, with a maximum at $\Delta\omega\neq 0$. The shift of  
the maximum of the resonant condition is associated with  the contribution  
from the nonlinear terms of the Hamiltonian (\ref{eq:Hvarphi}). Second,   
the maximual value of $n_0$ depends on the number of atoms transferred  
between the coupled states, which, in turn, depends on the coupling time  
$t_1$. This is related to the halfwidth $\Gamma\approx\frac{1}{t_1}$. If  
we compare both black lines in figure~\ref{fig:ramvsrabi}, we note that
$\Gamma_{\pi}\sim 1.44$ and $\Gamma_{\pi/2}\sim 2.76\approx
2\Gamma_{\pi}$.

The possibility of the formation of Ramsey fringes was first suggested in 
Ref.~\cite{Yukalov02}. Our numerical calculations demonstrate that 
the fringes are really obtained when two $\pi/2$ pulses are applied  
separately. In this way, given a fixed number of condensed atoms $N$,  
approximately half of the population is transferred from the ground to an 
excited state during the first $\pi/2$ pulse, evolving freely when the  
coupling is switched off, and then, a second pulse concludes the 
excitation. The fractional population after this process, as a function  
of the detuning $\Delta\omega$, is plotted in figure~\ref{fig:ramvsrabi} 
with the grey line. The process simulates the effect of the application  
of two $\pi/2$ Rabi pulses separated by $\tau=8t_1$. The maximal value  
for $n_p$ is found at the same frequency as in the configuration of a  
single pulse, which confirms our previous conclusion that the shift in the  
frequency is a hallmark of the nonlinear effects due to elastic two-body  
collisions.

\begin{figure}[ht]
%\vspace{0.5cm}
\centerline{\includegraphics[scale=1]{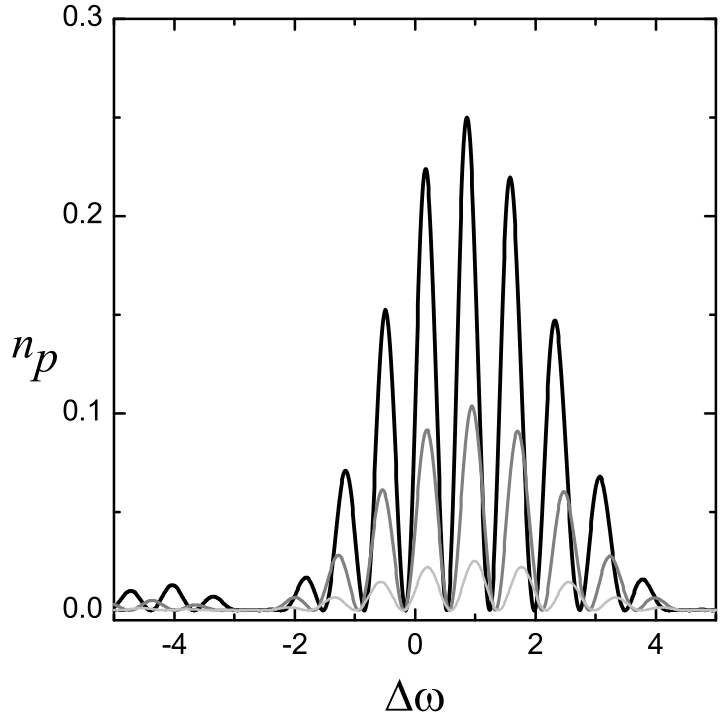}} \caption{Ramsey
fringes of the fractional populatuion $n_p$ as a function of  
$\Delta\omega$, after the application of two $\pi/2$- pulses separated by 
$\tau=4t_1$, with $t_1$ being the time needed for the excitation of 
the half of the population from the ground to a non-ground state. The 
lines correspond to the increasing values of $\beta$. Light grey line: 
$\beta=0.1\alpha$; Grey line: $\beta=0.2\alpha$; Black line:
$\beta=0.3\alpha$} \label{fig:ramvsd}
\end{figure}
\begin{figure}[ht]
%\vspace{0.5cm}
\centerline{\includegraphics[scale=1]{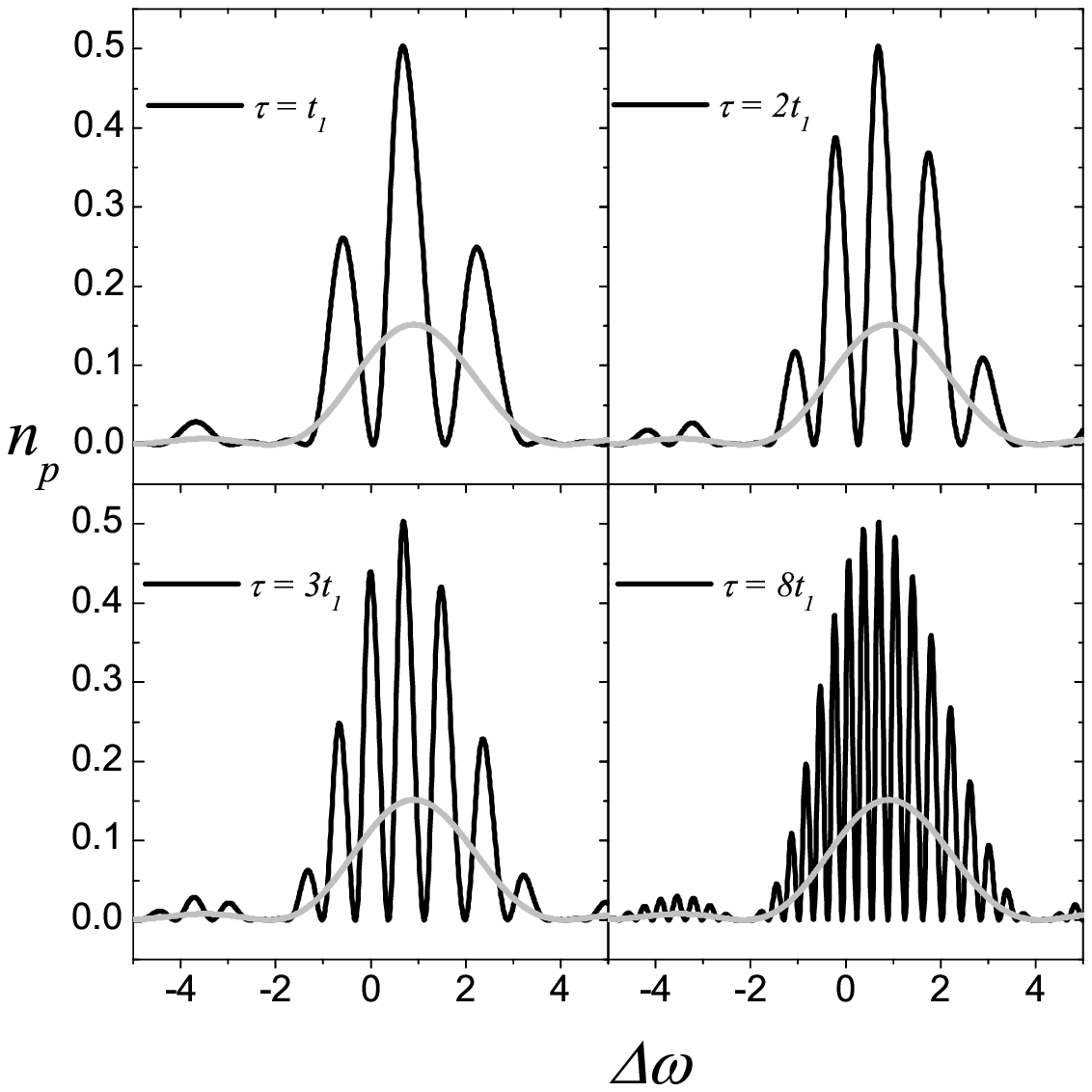}} \caption{Ramsey
fringes of the fractional population $n_p$ as a function of  
$\Delta\omega$, after the application of the Ramsey pulse configuration  
for different $\tau$ values. The gray line in all plots corresponds to the  
Rabi $\pi/2$-pulse.} \label{fig:ramtau}
\end{figure}

In figure~\ref{fig:ramvsd}, we plot the behavior of the fractional 
population as a function of $\beta$, considering the same pulse  
configuration as above, with an initial $\pi/2$ pulse, an interval   
$\tau=4t_1$, and a second $\pi/2$ pulse. If the coupling amplitude is  
weak, the Ramsey pattern loses its visibility. Increasing $\beta$ 
improves visibility, but at the same time causes power shifts,  
represented by the fringes displacement. The number of the Ramsey fringes,    
contained within the spectral width, is strongly dependent on the time 
interval $\tau$ between the applied $\pi/2$- pulses. This feature is  
shown in figure~\ref{fig:ramtau}, where four different situations are 
considered. We observe that the absolute maximal values of $n_p$ and its 
position, as a function of $\Delta\omega$, are not connected with 
the changes of $\tau$. What does depend on $\tau$ is the number 
of fringes, which increases as $\tau$ increases. If $\tau$ is 
equal to the coupling time $t_1$, two additional peaks appear
at the both sides of the main peak. For $\tau=2t_1$, we obtain two peaks 
at each side, and so on, as it can be seen comparing the four plots 
in figure~\ref{fig:ramtau}. The appearance of these auxiliary peaks is 
the hallmark for the accumulation of a relative phase between both 
topological states. During the time $\tau$, when the coupling is 
switched off, the dynamics of the whole system (ground + non-ground
state) is associated with the non-linear term in the Hamiltonian
(\ref{eq:Hvarphi}). This free dynamics determines the accumulated relative  
phase.

\section{Conclusions}
\label{sec:conclusions} In this work, we use numerical calculations 
for solving the system of coupled equations describing the resonant  
excitation of two coherent non-linear modes of BEC. Our results, obtained  
by means of the fourth-order Runge-Kutta method, give us an insight into  
the behavior of the fractional mode populations of non-ground collective  
atomic states in a harmonic trap. The principal novelty of the present  
paper is the investigation of the system response to different coupling  
configurations and the demonstration of the appearance of the Ramsey-like  
fringes.

The formation of the Ramsey-like fringes is a signature of the actual  
coherent character of topological modes in the studied nonlinear system.  
In both, Rabi and Ramsey pulse configurations, there is a shift of the  
maximum population transfer due to the strong effect of non-linearity of 
the system. The Ramsey pattern itself contains information on the 
accumulated relative phase, and the number of secondary peaks is 
proportional to the time $\tau$, as defined in Section~\ref{sec:Ramsey}.

As possible extension of this work, it would be interesting to 
consider the influence on the Ramsay fringes of the trap geometry and of  
different external pumping fields $V_p\deprt$. Other possibilities could  
be related to the manipilation of the scattering length \textit{via} 
the Feshbach resonance techniques and to the effects of changing the total  
number of atoms, which affects the interatomic intensity $\alpha$.

\section*{Acknowledgments} The authors wish to thank E.~P.~Yukalova for 
her important contribution at the early stage of this work. Special  
thanks are to E.~A.~L. Henn and K.~M.~F.~Magalhães for helpful 
discussions. This work was supported by Fapesp (Fundação de Amparo   
à Pesquisa do Estado de São Paulo), CAPES (Coordenação de Aperfeiçoamento 
de Pessoal de Nível Superior), and by CNPq (Conselho Nacional de  
Desenvolvimento Científico e Tecnológico).

\end{document}